# Factoring Odd Integers without Multiplication and Division

Charles Sauerbier

## 1 Introduction

Factoring of integers is a problem with a long history. The Sieve of Eratosthenes is perhaps the oldest know method. A number of methods[1] for factoring of integer have since been developed.

A method is interesting in the absence of need for multiplication and division in iterative component of the algorithm is presented. It can be used to determine if an integer is a prime, as the computation will not encounter a halting state at other than parameters that produce the original integer and 1. The method arose from empirical observation and reasoning of relations from what was observed, as opposed to theorizing a means on basis of prior knowledge or approach used by others.

This paper takes a less than conventional approach to presentation of the material. It is assumed the reader has familiarity with the underlying basic mathematics. The provided derivation follows from basic mathematical principles, so no proofs are provided.

## 2 Diophantine Expression

It is a property of integers that given some integer $n$ there exists two integers, $p$ and $q$, such that $n = p * q$, where $0 < p \leq n$ and $0 < q \leq n$. If $n$ is even then one solution is for $p = 2$ and $q = n/2$. Similarly should $q$ be an even integer then $q$ has 2 as a factor.

What complicates factoring is determining values for $p$ and $q$ where $n$ is an odd integer. The odd integers are not all multiples of a single integer value, as is the case for even integers. However, every odd integer $x$ can be expressed as $x = 2y + 1$, for some $y$ such that $0 \leq y \leq (x-1)/2$.

Considering only the case of factoring odd integers the set of equations in [2.1] is obtained.

| [2.1] | $n = p * q$ |
|---|---|
| | $n = 2a + 1$ |
| | $p = 2b + 1$ |
| | $q = 2c + 1$ |

The equations for $p$ and $q$ are consequence of the set of odd integers being closed under multiplication.

Using the equations of [2.1] a linear Diophantine equation in two unknowns, $b$ and $c$, is obtained. The derivation is in [2.2].

---

[1] See [2]

 

[2.2]
$$n = 2a + 1 = (2b + 1) * (2c + 1)$$
$$2a + 1 = 4bc + 2b + 2c + 1$$
$$2a = 4bc + 2b + 2c$$
$$a = 2bc + b + c$$
$$0 = 2bc + b + c - a$$

Solved for the integer roots the last expression in [2.2] allows the determination of the factors of $n$ where $b$ and $c$ are substituted back into the respective equations for $p$ and $q$ in [2.1].

*Conjecture 2.1*

The upper bound on complexity for determining the integer roots of $0 = 2bc + b + c - a$ is of the order $O\big((\log_2 n)^h\big)$, where $h$ is not dependent on $n$.

## 3 Difference Expressions

Given two initial values for $b$ and $c$ it is possible to adjust the values of each to approximate $a$. A system of equations to allow iterative approximation of [2.2] is presented in [2.3]. Iteration stops where $y_k = 0$.

[2.3]
$$a = (n - 1)/2$$
$$b_0 = c_0 = \left\lfloor \sqrt{a/2} \right\rfloor$$
$$y_k = |(2b_k c_k + b_k + c_k) - a|$$
$$b_k = \begin{cases} b_{k-1}, & 2b_{k-1} \leq y_{k-1} \\ b_{k-1} - 1, & 2b_{k-1} > y_{k-1} \end{cases}$$
$$c_k = c_k + 1$$

## 4 Removing Multiplication

It is possible to remove the multiplication operation in the iteration process. Transformation of the expression for $y_k$ for case where $2b_{k-1} < |y_{k-1}|$ is shown in [2.4], with case where $2b_{k-1} \geq |y_{k-1}|$ shown in [2.5].

 

| [2.4] | $b_k = b_{k-1}$ |
|---|---|
| | $c_k = c_{k-1} + 1$ |
| | $y_k = (2b_k c_k + b_k + c_k)$ |
| | $y_k = (2b_{k-1}(c_{k-1} + 1) + b_{k-1} + (c_{k-1} + 1))$ |
| | $y_k = (2b_{k-1}c_{k-1} + 2b_{k-1} + b_{k-1} + c_{k-1} + 1)$ |
| | $y_k = (2b_{k-1}c_{k-1} + b_{k-1} + c_{k-1}) + (2b_{k-1} + 1)$ |
| | $y_k = y_{k-1} + (2b_{k-1} + 1)$ |
| | $y_k = y_{k-1} + (2b_k + 1)$ |
| | $y_k = y_{k-1} + (b_k + b_k + 1)$ |

| [2.5] | $b_k = b_{k-1} - 1$ |
|---|---|
| | $c_k = c_{k-1} + 1$ |
| | $y_k = (2b_k c_k + b_k + c_k)$ |
| | $y_k = (2(b_{k-1} - 1)(c_{k-1} + 1) + (b_{k-1} - 1) + (c_{k-1} + 1))$ |
| | $y_k = 2(b_{k-1}c_{k-1} + b_{k-1} - c_{k-1} - 1) + b_{k-1} + c_{k-1}$ |
| | $y_k = 2b_{k-1}c_{k-1} + 2b_{k-1} - 2c_{k-1} - 2 + b_{k-1} + c_{k-1}$ |
| | $y_k = (2b_{k-1}c_{k-1} + b_{k-1} + c_{k-1}) + (2b_{k-1} - 2c_{k-1} - 2)$ |
| | $y_k = y_{k-1} + 2((b_k + 1) - (c_k - 1) - 1)$ |
| | $y_k = y_{k-1} + 2(b_k + 1 - c_k + 1 - 1)$ |
| | $y_k = y_{k-1} + 2(b_k - c_k + 1)$ |

The last expression in [2.4] and [2.5], respectively, allows the elimination of multiplication by reduction to simple addition of respective values. The expression for $y_k$, using the last expression of [2.4] and [2.5], respectively, is presented in [2.6].

| [2.6] | $y_0 = |(2b_0 c_0 + b_0 + c_0) - a|$ |
|---|---|
| | $y_k = \begin{cases} y_{k-1} + (2b_k + 1), & 2b_{k-1} < |y_{k-1}| \\ y_{k-1} + 2(b_k - c_k + 1), & 2b_{k-1} \geq |y_{k-1}| \end{cases}$ |

The set of computations in [2.3] are restated in [2.7] modified to include [2.6].

| [2.7] | $a = (n-1)/2$ |
|---|---|
| | $b_0 = c_0 = \lfloor\sqrt{x/2}\rfloor$ |
| | $y_0 = |(2b_0 c_0 + b_0 + c_0) - a|$ |
| | $y_k = \begin{cases} y_{k-1} + (2b_k + 1), & 2b_{k-1} < |y_{k-1}| \\ y_{k-1} + 2(b_k - c_k + 1), & 2b_{k-1} \geq |y_{k-1}| \end{cases}$ |
| | $b_k = \begin{cases} b_{k-1}, & 2b_{k-1} < |y_{k-1}| \\ b_{k-1} - 1, & 2b_{k-1} \geq |y_{k-1}| \end{cases}$ |
| | $c_k = c_k + 1$ |

The equations of [2.7] provide means to determine factors of an integer $n$ by iteration using only integer arithmetic operations of addition and subtraction. The computations were implemented using C#. The code for which is included in appendices. Empirical values used for testing were obtained using pairs of primes obtained at [1], with a sampling of values tested included in appendices.

## 5 Execution Time

Execution is limited by the monotonic decreasing sequence $b_0 \ldots b_k$. Though a monotonic decreasing sequence, it is not a strictly decreasing sequence. If for all iterations in execution $b_i \neq b_{i-1}$ the number of iterations is limited by $b_0$. Maximum execution time is therefore determined by both $b_0 \ldots b_k$ and number of instances where $b_i = b_{i-1}$.

Empirical observation of performance using product of pairs of primes obtained at [1] leads to the following conjecture:

*Conjecture 5.1*

> The limit on the value of $k$ in [2.7] in computation of factors of the product of two primes is $O\big((\log_2 n)^h\big)$, where $h$ is a function of $\log_2 q - \log_2 p$, where $p \leq q$.

## 6 Comments

The integer factoring method derived allows for computation to be performed using only integer addition and subtraction in iterative portion. While not presented, determination of initial values $b_0$ and $c_0$ can be determine without the need for square root operator. The division by 2 in determining value of $a$ can be accomplished in binary by shift operator, reducing the necessary operators to just three basic operators.

The implication of *Conjecture 5.1* impacts cryptographic methods that employ product of prime pairs. The difficulty of integer factoring using the method as presented is determined by the difference between integer factors.

    

## 8 Appendices
### 8.1 C# code

The C# code used in empirical testing is presented below.

```csharp
//****************************************
namespace IntFac
{
    using System.Numerics;
    //======================================
    class Program
    {
        //----------------------------------------
        static void Main()
        {
            System.Numerics.BigInteger[] factor_list = { 3, 31, 331, 3331, 33331, 333331,
                3333331, 33333331, 333333313, 982451653 };

            for (System.Int32 i = 0; i < factor_list.Length - 1; i++)
            {
                Factor(factor_list[i] * factor_list[i + 1]);
            }

            System.Console.ReadLine();
        }
        //----------------------------------------
        static void Factor(
            System.Numerics.BigInteger p_value_n)
        {
            System.Numerics.BigInteger val_n = p_value_n;
            System.Numerics.BigInteger val_p = 0;
            System.Numerics.BigInteger val_q = 0;
            System.Numerics.BigInteger val_a = 0;
            System.Numerics.BigInteger val_b = 0;
            System.Numerics.BigInteger val_c = 0;
            System.Numerics.BigInteger val_d = 0;
            System.Numerics.BigInteger val_e = 0;
            System.Numerics.BigInteger val_y = 0;

            System.Numerics.BigInteger tmp_d = 0;
            System.Numerics.BigInteger tmp_e = 0;

            val_a = (val_n - 1) / 2;
            val_b = val_c = (System.Numerics.BigInteger)System.Math.Sqrt((System.UInt64)val_a / 2);
            val_y = (2 * val_b * val_c) + val_b + val_c - val_a;

            System.Numerics.BigInteger i = 0;
            System.Console.Write(System.String.Format("{0},{1},", val_n, val_a));

            while ((val_b > 0) && (val_y != 0))
            {
                val_c += 1;
```

```csharp
                    val_d = val_b + val_b;
                    if (System.Numerics.BigInteger.Abs(val_y) > val_d)
                    {
                        val_y += (val_d + 1);
                        tmp_d = val_d + 1;
                    }
                    else
                    {
                        val_b -= 1;
                        val_e = val_b - val_c + 1;
                        val_y += (val_e + val_e);
                        tmp_e = val_e + val_e;
                    }
                    i += 1;
                    //
                    tmp_d = 0;
                    tmp_e = 0;
                }
                //
                System.Console.Write(System.String.Format("{0},{1},", val_b, val_c));
                //
                val_p = (val_b + val_b) + 1;
                val_q = (val_c + val_c) + 1;
                //
                System.Console.WriteLine(System.String.Format("{0},{1},{2},{3}",
                    val_p, val_q, val_p * val_q, i));
                
            }
            //-------------------------------------
        }   //  class
    }   //  namespace
```

## 8.2 Empirical Results

A sampling of results from empirical tests of method is presented in the table below.

| n | a | b | c | p | q | n | i |
|---|---|---|---|---|---|---|---|
| 96 | 46 | 1 | 15 | 3 | 31 | 93 | 11 |
| 10261 | 5130 | 15 | 165 | 31 | 331 | 10261 | 115 |
| 1102561 | 551280 | 165 | 1665 | 331 | 3331 | 1102561 | 1140 |
| 111025561 | 55512780 | 1665 | 16665 | 3331 | 33331 | 111025561 | 11397 |
| 11110255561 | 5555127780 | 16665 | 166665 | 33331 | 333331 | 11110255561 | 113963 |
| 1111102555561 | 555551277780 | 166665 | 1666665 | 333331 | 3333331 | 1111102555561 | 1139621 |
| 111111025555561 | 55555512777780 | 1666665 | 16666665 | 3333331 | 33333331 | 111111025555561 | 11396205 |
| 11111109655555603 | 5555554827777801 | 16666665 | 166666656 | 33333331 | 333333313 | 11111109655555603 | 113962032 |
| 327483864356816389 | 163741932178408194 | 166666656 | 491225826 | 333333313 | 982451653 | 327483864356816389 | 205094797 |
| 3333333133 | 1666666566 | 6 | 128205120 | 13 | 256410241 | 3333333133 | 128176253 |